\documentstyle[twoside,fleqn,nuclphys]{article}
\newcommand{\AmS}{{\protect\the\textfont2
  A\kern-.1667em\lower.5ex\hbox{M}\kern-.125emS}}

\title{Lepton--Quark Masses and Democratic Symmetry\thanks{Supported in part
                by DFG--contract 412/22--1 and EEC contract SC1--CT91--0729}
                \thanks{Invited talk given at the 3rd Workshop on Tau Lepton
                Physics, September 1994, Montreux, Switzerland}
                \thanks{Preprint--no.: MPI--PhT/94--77}}

\author{H. Fritzsch\address{Sektion Physik der Universit\"at M\"unchen,\\
        Theresienstrasse 37, D--80333 M\"unchen}
        \address{Max--Planck--Institut f\"ur Physik,
        Werner--Heisenberg-Institut,\\
        F\"ohringer Ring 6, D--80805 M\"unchen}
        }
\begin{document}
\begin{abstract}
It is shown that the simplest breaking of the subnuclear democracy leads
to
a successful description of the mixing between the second and third
family.
In the lepton channel the $\nu _{\mu } - \nu_{\tau }$ oscillations are
expected to be described by a mixing angle of $2.65^ {\circ }$ which
might
be observed soon in neutrino experiments.
\end{abstract}

\maketitle
\setcounter{page}{1}
In the standard electroweak model both the masses of the quarks as well
as
the weak mixing angles enter as free parameters. Any further insight
into
the yet unknown dynamics of mass generation would imply a step beyond
the
 physics
of the electroweak standard model. At present it seems far too early to
attempt an actual solution of the dynamics of mass generation, and one
is
invited to follow a strategy similar to the one which led eventually to
the
solution of the strong interaction dynamics by QCD, by looking for
specific patterns and symmetries as well as specific symmetry
violations.\\
\\
\indent
The mass spectra of the quarks are dominated essentially by the masses of
the
members of the third family, i.\ e.\ by $t$ and $b$. Thus a clear
hierarchical
pattern exists. Furthermore the masses of the first family are small
compared
to those of the second one. Moreover, the CKM--mixing matrix exhibits a
hierarchical pattern -- the transitions between the second and third
family
as well as between the first and the third family are small compared to
those between the first and the second family.\\
\\
\indent
About 15 years ago, it was emphasized$^{1)}$ that the observed
hierarchies signify
that nature seems to be close to the so--called ``rank--one'' limit, in
which
all mixing angles vanish and both the u-- and d--type mass matrices are
proportional to the rank-one matrix
\begin{equation}
M_0 = {\rm const.} \cdot \left(\begin{array}{ccc}
0 & 0 & 0\\ 0 & 0 & 0\\ 0 & 0 & 1 \end{array} \right) \, .
\end{equation}
\hspace*{0.1cm}
Whether the dynamics of the mass generation allows that this limit can
be
achieved in a consistent way remains an unsolved issue, depending on the
dynamical details of mass generation. Encouraged by
the
observed hierarchical pattern of the masses and the mixing parameters,
we
shall assume that this is the case. In itself it is a non-trivial
constraint
and can be derived from imposing a chiral symmetry, as emphasized in
ref. (2).
This symmetry ensures that an electroweak doublet which is massless
remains
unmixed and is coupled to the $W$--boson with full strength.\\
\\
\indent
As soon as
the mass is introduced, at least for one member of the doublet, the symmetry
is
violated and mixing phenomena are expected to show up. That way a chiral
evolution of the CKM matrix can be constructed.$^{2)}$ At the first stage
only the $t$ and $b$ quark masses are introduced, due to their
non-vanishing
coupling to the scalar ``Higgs'' field. The CKM--matrix is unity in this
limit. At the next stage the second generation acquires a mass.
Since
the $(u, d)$--doublet is still massless, only the second and the third
 generations
mix, and the CKM--matrix is given by a real $2 \times 2$ rotation matrix
in the
$(c, \, s) - (t, \, b)$ subsystem, describing e.\ g.\ the mixing between
$s$
and $b$. Only at the next step, at which the $u$ and $d$ masses are
introduced, does the full CKM--matrix appear, described in general by
three angles
 and one
phase.\\
\\
\indent
It has been emphasized some time ago$^{4, \, 5)}$ that the rank-one mass
matrix (see
 eq.
(1)) can be expressed in terms of a ``democratic mass matrix'':
\begin{equation}
M_0 = c \left( \begin{array}{ccc}
1 & 1 & 1\\ 1 & 1 & 1\\ 1 & 1 & 1 \end{array} \right) \, ,
\end{equation}
which exhibits an $S(3)_L \, \, \times \, \, S(3)_R$ symmetry. Writing
down
the mass eigenstates in terms of the eigenstates of the
``democratic'' symmetry, one finds e.g. for the lepton channel:
\begin{eqnarray}
e^0 & = & \frac{1}{\sqrt{2}} (l_1 - l_2) \nonumber\\
\mu^0 & = & \frac{1}{\sqrt{6}} (l_1 + l_2 - 2l_3)\\
\tau^0 & = & \frac{1}{\sqrt{3}} (l_1 + l_2 + l_3) \nonumber .
\end{eqnarray}
Here $l_1, \ldots$ are the symmetry eigenstates.
Note that $e^0$ and $\mu^0$ are massless
in
the limit considered here, and any linear combination of the first two
state vectors given in eq. (3) would fulfill the same purpose, i.\ e.\
the
decomposition is not unique, only the wave function of the coherent
state
$\tau^0$ is uniquely defined. This ambiguity will disappear as soon as
the symmetry is violated.\\
\\
\indent
The wave functions given in eq. (3) are reminiscent of the wave
functions
of the neutral pseudoscalar mesons in QCD in the $SU(3)_L \, \, \times
\, \,
SU(3)_R$ limit:
\begin{eqnarray}
\pi^0_0 & = & \frac{1}{\sqrt{2}}(\bar uu - \bar dd)\\
\eta_0 & = & \frac{1}{\sqrt{6}}(\bar uu + \bar dd - 2\bar ss)
\nonumber\\
\eta '_0 & = & \frac{1}{\sqrt{3}}(\bar uu + \bar dd + \bar ss) .
\nonumber
\end{eqnarray}
(Here the lower index denotes that we are considering the chiral limit).
Also the mass spectrum of these mesons is identical to the mass spectrum
of
the leptons and quarks in the ``democratic'' limit: two mesons $(\pi
^0_0 \, ,
\eta _0)$ are massless and act as Nambu--Goldstone bosons, while the
third
coherent state $\eta '_0$ is \underline{not} massless due to the QCD
anomaly.\\
\\
\indent
In the chiral limit the (mass)$^2$--matrix of the neutral pseudoscalar
mesons
is also a ``democratic'' mass matrix when written in terms of the $(\bar
qq)$--
eigenstates $(\bar uu), \, (\bar dd)$ and $(\bar ss)^{5)}$:
\begin{equation}
M^2(ps) = \lambda \left( \begin{array}{ccc}
1 & 1 & 1\\ 1 & 1 & 1\\ 1 & 1 & 1 \end{array} \right),
\end{equation}
where the strength parameter $\lambda $ is given by
$\lambda = M^{2}(\eta '_{0}) \, / \, 3$.
The mass matrix (5) describes the result of the QCD--anomaly which
causes strong
transitions between the quark eigenstates (due to gluonic annihilation
effects
enhanced by topological effects). Likewise one may argue that analogous
transitions are the reason for
the lepton--quark mass hierarchy. Here we shall not speculate about a
detailed mechanism of this type, but merely study the effect of symmetry
breaking.\\
\\
\indent
In the case of the pseudoscalar mesons the breaking of
the symmetry down to
$SU(2)_L \, \, \times \, \, SU(2)_R$ is provided by a direct mass term
$m_s \bar
 ss$
for the s--quark. This implies a modifica\-tion of the (3,3) matrix
element in
eq. (5), where $\lambda $ is replaced by $\lambda + M^2(\bar ss)$ where
 $M^2(\bar ss)$ is
given by $2M^2_k$, which is proportional to $< \bar ss >_0$, the
expectation
value of $\bar ss$ in the QCD vacuum. This direct mass term causes the
violation of the symmetry and generates at the same time a mixing
between
$\eta _0$ and $\eta '_{0}$, a mass for the $\eta _{0}$, and a mass shift
for
the $\eta '_{0}$.\\
\\
\indent
It would be interesting to see whether an analogue of the
simplest violation of the ``democratic'' symmetry which describes successfully
the mass and mixing pattern of the $\eta - \eta '$--system is also able to
describe the observed mixing and mass pattern of the second and third
family of leptons and quarks. This was discussed recently$^{6)}$.
Let us replace the (3,3) matrix element in
eq. (2) by $1 + \varepsilon _i$; (i = l (leptons), u (u--quarks), d
(d--quarks)
respectively. The small real parameters $\varepsilon _i$ describe the
departure
from democratic symmetry and lead
\begin{enumerate}
\item[a)]
to a generation of mass for the second family and
\item[b)] to a flavour mixing between the third and the second
family. Since $\varepsilon $ is directly related (see below) to a
fermion mass
and the latter is \underline{not} restricted to be positive,
$\varepsilon $
can be positive or negative. (Note that a negative Fermi--Dirac mass can
always be turned into a positive one by a suitable $\gamma
_5$--transformation
of the spin $\frac{1}{2}$ field). Since the original mass term is
represented by a symmetric matrix, we take $\varepsilon $ to be real.
\end{enumerate}
\hspace*{0.1cm}
In ref. (4) a general breaking of the flavor democracy was discussed in
term of two parameters $\alpha $ and $\beta $. The ansatz diskussed here,
in analogy to the case of the pseudoscalar mesons which represents the
simplest breaking of the flavor democracy, corresponds to the special case
$\alpha = 0$. Note that the case $\beta = \alpha + \alpha^*$
discussed in ref. (11) leads to the mass matrix given in ref. (1).\\
\\
\indent
First we study the mass and mixing pattern of the charged
leptons. The mass
operator (trace $\Theta^{\mu} _{\mu}$ of the energy--momentum tensor
$\Theta_{\mu \nu})$ can be written as
\begin{equation}
\Theta ^{\mu }_{\mu } = \Theta ^{0 \mu }_{\mu } + c_l \varepsilon _l
\bar l_3 l_3
\end{equation}
where $\Theta ^{0 \mu}_\mu $ describes the mass term in the symmetry
limit.
The modification of the spectrum and the induced mixing can be obtained
by
considering the matrix elements:
\begin{eqnarray}
< \mu ^0 | c_l \varepsilon _l \bar l_3 l_3 | \mu ^0 > & = & +
\frac{2}{3} c_l
\varepsilon _l \nonumber\\
< \tau ^0 | c_l \varepsilon _l \bar l_3 l_3 | \tau ^0 > & = & +
\frac{1}{3}
c_l \varepsilon _l\\
< \mu ^0 | c_l \varepsilon _l \bar l_3 l_3 | \tau ^0 > & = & -
\frac{\sqrt{2}}
{3} c_l \varepsilon _l \nonumber \, \, .
\end{eqnarray}
\hspace*{0.1cm}
One observes that
\begin{enumerate}
\item[a)] the muon acquires a mass given by
$c_l
\cdot \varepsilon _l$ i.\ e.\ $m(\mu ) / m(\tau ) \cong \frac{2}{9}
\varepsilon _l$;
\item[b)] the $\tau $--lepton mass is changed
slightly
($m(\tau ) / m(\tau ^0) \cong 1 + \frac{1}{9} \varepsilon _l)$;
\item[c)] the flavour mixing is induced -
the perturbation proportional to $\bar l _3 l_3$ leads to a
non--vanishing
transition matrix element between $\mu ^0$ und $\tau ^0$.
\end{enumerate}
\hspace*{0.1cm}
This phenomenon is
analogous to the chiral symmetry violation of QCD, where the s--quark
mass
term $m_s \bar ss$ leads to a mass for the $\eta $--meson, a mass shift
for
the $\eta '$--meson and a mixing between $\eta $ and $\eta '$.\\
\\
\indent
It is instructive to rewrite the mass matrix in the hierarchical basis,
where
one obtains, using the relations (7):
\begin{equation}
M = c_l \left( \begin{array}{ccc}
0 & 0 & 0\\ 0 & + \frac{2}{3}{\varepsilon _l} & - \frac{\sqrt{2}}{3}
\varepsilon
 _l\\
0 & - \frac{\sqrt{2}}{3} \varepsilon _l & 3 + \frac{1}{3}\varepsilon _l
 \end{array}
\right) \, .
\end{equation}
\hspace*{0.1cm}
In lowest order of $\varepsilon $ one finds the mass eigenvalues
$m_\mu = \frac{2}{9} \varepsilon _l \cdot m_\tau \, , m_\tau = m_{\tau
^0} \, ,
\Theta _{\mu \tau} = | \sqrt{2} \cdot \varepsilon _l / 9|$.\\
\\
\indent
The exact mass eigenvalues and the mixing angle are given by:
\begin{eqnarray}
m_1 / c_l & = & \frac{3 + \varepsilon _l}{2} - \frac{3}{2} \sqrt{1 -
\frac{2}{9} \varepsilon _l + \frac{1}{9} \varepsilon _l ^2} \nonumber\\
m_2 / c_l & = & \frac{3 + \varepsilon _l}{2} + \frac{3}{2} \sqrt{1 -
\frac{2}{9}
\varepsilon _l + \frac{1}{9} \varepsilon _l^2}\\
\sin \Theta _l & = & \frac{1}{\sqrt{2}}\left( 1 - \frac{1 -
 \frac{1}{9}\varepsilon _l}
{(1 - \frac{2}{9} \varepsilon _l + \frac{1}{9} \varepsilon
 ^2_l)^{1/2}}\right)^{1/2} \nonumber
\end{eqnarray}
\hspace*{0.1cm}
The ratio $m_{\mu } / m_{\tau }$, observed to be $0.0595$, gives
$\varepsilon _l = 0.286$ and a $\mu - \tau $ mixing angle of
$2.65^{\circ}$.
Whether this mixing angle is directly relevant for neutrino oscillations
or
not depends on the neutrino sector. For massless neutrinos the mixing
angle
does not have a direct physical meaning, i.\ e.\ it can be rotated away.
If
neutrinos have a mass, the neutrino mass matrix will in general induce
further
mixing angles.\\
\\
\indent
A general discussion would be beyond the scope of this
paper.
However, we should like to consider an interesting scenario which is
being
discussed in connection with cosmology aspects. Let us suppose that
the $\tau $--neutrino mass is of the order of 10 eV in order to be
relevant for
the ``missing matter problem'' in cosmology.
Taking into account possible hints towards neutrino oscillations from the
solar neutrino experiments, one concludes$^{7)}$
$10^{-4} \leq m(\nu_{\tau }) \leq 10^{-3}$. Under the assumption that neutrinos
are Dirac particles like
the charged leptons and that our considerations about the democratic symmetry
and its breaking are applied both for them and the charged leptons, we
conclude: The $\varepsilon $--parameter for the neutrino sector is tiny
$(< 5 \cdot 10^{-3})$,
and the mixing angle induced via the $\nu _{\mu }$--mass generation can
safely be neglected. Thus the angle relevant for the
$\nu _{\mu } - \nu _{\tau }$ oscillations remains
$2.65^{\circ }$, i.\ e.\ $\sin^{2} 2\Theta = 0.0085$.
This value is essentially the lowest limit given by the Charm II
experiment$^{8)}$,
i.\ e.\ is not ruled out for any value of
$\bigtriangleup m^2 = m(\nu _{\tau })^2 - m(\nu _{\mu })^2$.
However, the E(531) experiment$^{9)}$ gives a limit of about $16 eV^2$ for
$\bigtriangleup m^2$, i.\ e.\ $m(\nu _{\tau }) < 4 eV$. This limit seems
to
rule out a cosmological role with respect to the ``missing matter'' for
the
$\tau $--neutrino. However, one might caution this conclusion since our
mixing angle of $2.65^{\circ }$ is not far from the limit of
$(\sin^2 2 \Theta = 0.004)$, at which, according to the E531 experiment,
all
values of $m(\nu _{\tau })$ are allowed. New experiments, e.\ g.\ the
CHORUS and NOMAD experiments now or soon under way at CERN, will clarify
this issue.
If the mixing angle is $2.65^{\circ }$ as argued above and the $\nu
_{\tau }$--mass
above 10 eV, one should observe the $\nu _{\mu } - \nu _{\tau }$--
oscillations within one year$^{9)}$.\\
\\
\indent
Replacing $\varepsilon _l$ by $\varepsilon _u$, $\varepsilon _d$
respectively, we can determine the symmetry breaking parameters for the
quark sector. The ratio $m_s / m_b$ is allowed to vary in the range
$0.022 \ldots 0.044$ (see ref. (10)). According to eq. (9) one finds
$\varepsilon _u$ to vary from $\varepsilon _d = 0.11$ to $0.21$.
The associated $s - b$
mixing angle varies from $\Theta (s, b) = 1.0^{\circ }$ \hspace{0.3cm}
$(\sin \Theta = 0.018)$ and $\Theta (s, b) = 1.95^{\circ }$
\hspace{0.3cm}
$(\sin \Theta = 0.034)$. As an illustrative example we use the values
$m_b(1$GeV$)
 = 5200 $MeV, \hspace{0.3cm}
$m_s(1$GeV$) = 220 $MeV. One obtains $\varepsilon _d = 0.20$ and
$\sin \Theta(s, b) = 0.032$.\\
\\
\indent
To determine the amount of mixing in the $(c, t)$--channel, a knowledge
of
the ratio $m_c / m_t$ is required. As an illustrative example we take
$m_c(1$GeV$) = 1.35 $GeV, $m_t(1GeV) = 260 $GeV (i.\ e.\ $m_t(m_t) =
160 $GeV$)$,
which gives $m_c / m_t \cong 0.005$. In this case one finds $\varepsilon
_u =
 0.023$ and $\Theta(c,t) = 0.21^{\circ }$
\hspace{0.3cm} $(\sin \Theta (c, t) = 0.004$) .\\
\\
\indent
The actual weak mixing between the third and the second quark family is
combined effect of the two family mixings described above. The symmetry
breaking given by the $\varepsilon $--parameter can be interpreted, as
done
in eq. (7), as a direct mass term for the $l_3(u_3, d_3)$ fermion
system.
However, a direct fermion mass term need not be positive, since its sign
can always be changed by a suitable $\gamma _5$--transformation. What
counts
for our ana\-lysis is the relative sign of the $m_s$--mass term in
comparison
to the $m_c$--term, discussed previously. Thus two possibilities must be
considered:
\begin{enumerate}
\item[a)] Both the $m_s$-- and the $m_c$--term
have the same relative sign with respect to each other, i.\ e.\ both
 $\varepsilon _d$
and $\varepsilon _u$ are positive, and the mixing angle between the
second
and third family is given by the difference $\Theta (sb) - \Theta (ct)$.
This
possibility seems to be ruled out by experiment, since it would lead to
$V_{cb} < 0.03$.
\item[b)] The relative signs of the breaking
terms
$\varepsilon _d$ and $\varepsilon _u$ are different, and the mixing
angle
between the $(s,b)$ and $(c,t)$ systems is given by the sum
$\Theta(sb) + \Theta(ct)$. Thus we obtain $V_{cb} \cong \sin (\Theta(sb)
+ \Theta(ct))$.
\end{enumerate}
\hspace*{0.1cm}
According to the range of values for $m_s$ discussed above, one finds
$V_{cb} \cong 0.022 ... 0.038$.
For example, for $m_s(1$GeV$) = 220$MeV, \, $m_c (1$GeV$) = 1.35 $GeV,
$m_t(1$GeV$) = 260$GeV one finds $V_{cb} \cong 0.036$.\\
\\
\indent
Before discussing the experimental situation, we add a comment about the
mass
ge\-neration for the first family, which at the same time will also
generate
the other mixing elements, e.g. $V_{us}$ and $V_{ub}$, of the CKM
matrix. These
masses can be generated by a further breakdown of the symmetry, e.\ g.\
in the
matrix of eq. (5) by a small departure of a second diagonal matrix
element
from unity. (This would correspond to a direct mass term for that
state.) Due to the small values of the masses of the first family in
comparison to the $\lambda $--scale, given by the mass of the third
generation fermion (e.g. $m_e / \lambda = 0.0009)$, the strength of this
symmetry breaking is much smaller than the primary symmetry breaking,
which
leads to the masses for the second family. (The situation is analogous
to the
one in hadronic physics, where the breaking of the chiral symmetry is
primarily
given by the mass of the s--quark, and the $m_u / m_d$ mass terms can
be neglected to a good approximation). In general it is expected, both
from the
 arguments considered here and more
generally from the analysis on chiral symmetry given in ref. (2), that
the
matrix elements $V_{cb}$ and $V_{ts}$ will be affected only by small
corrections
of order $10^{-3}$ or less in absolute magnitude (of order
$\frac{m_d}{m_b}$,
$\frac{m_u}{m_t}$ respectively). Thus the primary breaking of the
democratic symmetry leads solely to a mixing between the second and the
third
family, and the secondary breaking, responsible for the Cabibbo angle
etc.,
will not affect the $2 \times 2$ submatrix of the CKM--matrix describing
the $s
 - b$
mixing in a significant way.\\
\\
\indent
The experiments give $V_{cb} = 0.032 \dots 0.054^{12)}$. We conclude from
the analysis
 given above
that our ansatz for the symmetry breaking reproduces the lower part of
the
experimental range. According to a recent analysis the experimental data
are reproduced best for $V_{cb} = 0.038 \pm 0.003^{12)}$, i.\ e.\ it
seems
that $V_{cb}$ is lower than previously thought, consistent with our
expectation. Nevertheless we obtain consistency with experiment only if
the ration $m_s / m_b$ is relatively large implying $m_s(1$GeV$) \ge
180$MeV.\\
\\
\indent
It is remarkable that the simplest ansatz for the breaking of the
``democratic
symmetry'', one which nature follows in the case of the pseudoscalar
mesons, is able to reproduce the experimental data on the mixing between
the second and third family. We interpret this as a hint that the
eigenstates
of the symmetry $l_{i}, q_i$ respectively, and not the mass eigenstates,
play
a special r\^{o}le in the physics of flavour, a r\^{o}le
which needs to be investigated further.\\
\indent
Finally we should like to add to comment with regard to the mass generation
for the first lepton--quark family. It is well--known that the ansatz
discussed in ref.\ (1), generalising earlier work of S. Weinberg$^{14)}$ and
one of the authors$^{15)}$, is able to describe the mixing and mass generation
for the first generation very well. Taking this into account, it is easy to
see that the following ansatz for the mass matrix generalising eq.\ (8) is able
to describe the mass and mixing pattern for all three
generations ($\delta $: second, complex parameter)$^{16}$:
\begin{equation}
M = c \cdot \left( \begin{array}{l}
                    0\\ \delta ^* \\ 0 \end{array}
                    \begin{array}{c}
                    \delta \\ \frac{2}{3} \varepsilon\\
                    - \frac{\sqrt{2}}{3} \varepsilon \end{array}
                    \begin{array}{c}
                    0\\ - \frac{\sqrt{2}}{3} \varepsilon \\
                    3 + \frac{1}{3} \varepsilon \end{array}
                    \right)
\end{equation}
\hspace*{0.1cm}
However, the breaking of democratic symmetry required to reproduce
eq.\ (10) is such that $\delta $ does not appear solely in the diagonal
elements, but in particular in the (1,3) and (2,3) elements of the
democratic mass matrix. A deeper understanding of this feature of the
symmetry breaking is still lacking.\\
\\
\indent
Alternatively one might try to reproduce the Cabibbo type mixing solely
by modifying the (1,1) and (2,2) diagonal elements. However in this case
it is difficult to obtain an $s$--quark mass small enough to be consistent
with observation$^{17)}$, and we conclude that at least
a small violation of the democratic symmetry in the off--diagonal elements
is needed in order to understand the mass and mixing pattern for the first
family. In view of the fact that the masses of the members of the first
family are tiny compared to the masses of the members of the third family,
we are not discouraged. Also in the case of the pseudoscalar mesons it is
known that off--diagonal breaking terms of the democratic symmetry are
required in order to understand the fine details of the spectrum$^{18)}$.
Further investigations in order to understand the mass spectrum of the
first family and CP--violation are in progress.\\
\newpage
\noindent
\underline{Discussion}\\
\\
M.\ Roney, I.P.P., University of Victoria:\\
Given the $\tau- \mu $--mixing in the model of democratic symmetry, at
what rate can one expect $\mu \rightarrow e \gamma $?\\
\\
Answer:\\
The $\tau- \mu$--mixing, as described here, does not imply that
a decay like $\mu \rightarrow e \gamma $ will show up. Since we are still
within the framework of the Standard Model, this decay would exist only,
if neutrinos are massive. However, neutrino masses in the eV--range would
induce a decay, whose branching ratio is negligible (less than $10^{-20}$).
Of course, the situation would change, if elements are added which go beyond
the Standard Model.\\
\\
S.\ Narison, Montpellier and CERN:\\
What is the role of the $U(1)$ anomaly in the $\mu- \tau$--splitting in
addition to the $SU(3)$ breaking effect? What is the effect of
$m_b $ or $m_s $ in the prediction of the model $(V_{cb} \ldots )$?\\
\\
Answer:\\
The $U(1)$ anomaly was used only in the analogy with the pseudoscalar
spectrum. In the approach based on the ``democratic symmetry'' the
symmetry limit is similar to the one induced by the anomaly for the
pseudoscalars. This does not imply that an anomaly is at work here.
However, the symmetry breaking is analogous. The $\mu- \tau$--splitting
corresponds to the $\eta - \eta '$--splitting. The ratio $m_s / m_b$
determines $V_{cb}$. Thus a relatively large value of $m_s$, around
200 MeV, and a small value of $m_b (1 $GEV$)$, say about 5.3 GeV, would be
welcome.\\
\\
N.\ Wermes, Bonn:\\
What kind of an effect could the $\mu- \tau$--mixing introduce for
$(g-2)_{\tau }$?\\
\\
Answer:\\
The effect would be negligible, also for the magnetic transition
$\tau \rightarrow \mu \gamma $, if we don't leave the area of the Standard
Model. However, if the typical energy range relevant for the ``democratic
symmetry'' is the 1 TeV range, interesting effects could be expected,
e.\ g.\ a sizeable fraction for the decay $\tau \rightarrow \mu \gamma $.\\
\\
A.\ Weinstein:\\
Given the quark masses, can you predict the Cabibbo angle, the KM--phase
and $V_{cb}$?\\
\\
Answer:\\
Using the simplest breaking of the democratic symmetry, one can calculate
$V_{cb}$ in terms of the mass ratios $m_s / m_b$ and $m_c / m_t$, in
agreement with experiment. Furthermore the specific model mentioned at the
end allows one to calculate the Cabibbo angle and the KM--phase, the
latter being close to 90 degrees. The Cabibbo angle is essentially given by
$sin \Theta_c \cong (m_d / m_s)^{1/2}$.\\
\\

\end{document}